\documentclass[notitlepage,twocolumn]{revtex4-1}
\usepackage[utf8]{inputenc}
\usepackage{amsmath}
\usepackage{amsfonts}
\usepackage{amssymb}
\usepackage{times,fullpage}
\usepackage{comment,subfigure}
\usepackage{array,graphicx}
\usepackage{color}

\graphicspath{{./figures/}}

\begin{document}

\newcommand{\nwc}{\newcommand}
\nwc{\vs}{\vspace}
\nwc{\hs}{\hspace}
\nwc{\la}{\langle}
\nwc{\ra}{\rangle}
\nwc{\nn}{\nonumber}
\nwc{\Ra}{\Rightarrow}
\nwc{\wt}{\widetilde}
\nwc{\lw}{\linewidth}
\nwc{\ft}{\frametitle}
\nwc{\ben}{\begin{enumerate}}
  \nwc{\een}{\end{enumerate}}
\nwc{\bit}{\begin{itemize}}
  \nwc{\eit}{\end{itemize}}
\nwc{\dg}{\dagger}
\nwc{\mA}{\mathcal A}
\nwc{\mD}{\mathcal D}
\nwc{\mB}{\mathcal B}

\nwc{\Tr}[1]{\underset{#1}{\mbox{Tr}}~}
\nwc{\pd}[2]{\frac{\partial #1}{\partial #2}}
\nwc{\ppd}[2]{\frac{\partial^2 #1}{\partial #2^2}}
\nwc{\fd}[2]{\frac{\delta #1}{\delta #2}}
\nwc{\pr}[2]{K(i_{#1},\alpha_{#1}|i_{#2},\alpha_{#2})}
\nwc{\av}[1]{\left< #1\right>}

\nwc{\zprl}[3]{Phys. Rev. Lett. ~{\bf #1},~#2~(#3)}
\nwc{\zpre}[3]{Phys. Rev. E ~{\bf #1},~#2~(#3)}
\nwc{\zpra}[3]{Phys. Rev. A ~{\bf #1},~#2~(#3)}
\nwc{\zjsm}[3]{J. Stat. Mech. ~{\bf #1},~#2~(#3)}
\nwc{\zepjb}[3]{Eur. Phys. J. B ~{\bf #1},~#2~(#3)}
\nwc{\zrmp}[3]{Rev. Mod. Phys. ~{\bf #1},~#2~(#3)}
\nwc{\zepl}[3]{Europhys. Lett. ~{\bf #1},~#2~(#3)}
\nwc{\zjsp}[3]{J. Stat. Phys. ~{\bf #1},~#2~(#3)}
\nwc{\zptps}[3]{Prog. Theor. Phys. Suppl. ~{\bf #1},~#2~(#3)}
\nwc{\zpt}[3]{Physics Today ~{\bf #1},~#2~(#3)}
\nwc{\zap}[3]{Adv. Phys. ~{\bf #1},~#2~(#3)}
\nwc{\zjpcm}[3]{J. Phys. Condens. Matter ~{\bf #1},~#2~(#3)}
\nwc{\zjpa}[3]{J. Phys. A ~{\bf #1},~#2~(#3)}
\nwc{\zpjp}[3]{Pramana J. Phys. ~{\bf #1},~#2~(#3)}

\title{Return to origin problem for particle on a one-dimensional lattice with quasi-Zeno dynamics}
\author{Sourabh Lahiri$^{1,2}$}
\email{sourabhlahiri@gmail.com}
\author{Abhishek Dhar$^1$}
\email{abhishek.dhar@icts.res.in}

\affiliation{$^1$International Centre for Theoretical Sciences, Tata Institute of Fundamental Research, Survey no. 151, Sivakote, Hesaraghatta Hobli, Bengaluru 560089, India\\
  $^2$ Department of Physics, Birla Institute of Technology Mesra, Ranchi 835215, Jharkhand, India.
}

\begin{abstract}
  In recent work, the so-called quasi-Zeno dynamics of a system has been investigated in the context of the quantum first passage problem. This dynamics considers the  time evolution of a system subjected to a sequence of selective projective measurements made at small but finite intervals of time.  This  means that one has  a sequence of steps, with each step consisting of a  
  unitary transformation followed by a projection. The dynamics is  non-unitary and, in recent work, it has been shown that it can be effectively described by two different  non-Hermitian Hamiltonians. Here we explore this connection by considering the problem of detecting a free quantum particle moving on a one-dimensional lattice,  where the detector is placed at the origin and the particle is initially located at some specified  lattice point. We find that results for distribution times for the first detection probability, obtained from the non-Hermitian Hamiltonians, are in excellent agreement with known exact results as well as exact numerics. Interesting finite-size effects are discussed. We also study the first detection problem for the example of a particle moving in a quasi-periodic potential, an example where the unperturbed particle's motion can be ballistic, localized or diffusive.
\end{abstract}

\maketitle

\section{Introduction}

We consider a quantum particle moving on a one-dimensional lattice with its dynamics described by a tight-binding Hamiltonian. The particle starts from some initial position, and is allowed to undergo nearest-neighbour hoppings. A detector is placed at a pre-determined site, say the origin. The evolving wave-function of the particle is subjected to a sequence of measurements by the detector after every $\tau$ time-interval. The experiment is stopped as soon as the particle is detected for the first time.
Thus, the unitary evolution of the particle  is interrupted by  a sequence of projective measurements that are made at the origin, at regular intervals of time $\tau$. This kind of dynamics has been studied in a number of recent papers \cite{dha15_pra,dha15_jpa,friedman2016quantum,friedman2017quantum} and,  for the case where $\tau$ is small but \emph{finite}, this  has been referred to as \emph{quasi-Zeno dynamics} \cite{elliott2016quantum}. We note that the limit $\tau \to 0$  corresponds to the case of continuous measurements on a system and leads to the famous so-called quantum Zeno effect \cite{misra1977zeno}, whereby the quantum particle will never be detected. This is easy to understand physically. Suppose we divide the full Hilbert space where the particle lives, into a part where the detection occurs, say $\mathcal{H}_D$ and the rest of the space is labeled $\mathcal{H}_S$.
If the particle is initially located in $\mathcal{H}_S$, then selective measurements essentially project the system  back into this space and, in the limit of continuous measurements, the system remains in this space. The Zeno effect was first observed experimentally in \cite{wineland90} but there has been some confusion regarding the interpretation of the experiment \cite{ballentine,wineland91,pascazio94}.  
  By \emph{not} taking the limit $\tau \to 0$, one uncovers a rich behavior of the dynamics and interesting physical questions can be asked. In particular, the particle now gets detected at a finite time and one can ask for the distribution of  detection times.

The papers \cite{dha15_pra,dha15_jpa} looked at a single free particle in finite lattices in one and higher dimensions and explored the question of first detection times and survival probability by localized detectors. Using perturbation theory it was shown that the dynamics of the system can be understood as  evolution under effective non-Hermitian Hamiltonians. Analytic results were obtained and compared to exact numerical results.
The authors in \cite{friedman2016quantum,friedman2017quantum,thiel2017first} developed a renewal time approach, commonly used in the area of classical random walks in the context of first passage time probabilities, to address the question of computing the first detection time (FDT) distribution in the asymptotic time limit. Using this approach they obtained exact results for the first detection time distribution of a particle on an infinite lattice, for the case where  the detector is placed at the origin. A more general discussion of quasi-Zeno dynamics, including in interacting systems, has been given in \cite{elliott2016quantum} while non-Hermitian Hamiltonians arising in the 
context of measurements have been discussed in \cite{halliwell2009,kozlowski2016non}. Finally, we note that a similar protocol of repeated measurements has been  used to define first detection probabilities for quantum random walks \cite{bach2004,krovi2006,kiss2008,grunbaum2013}.

In this paper we investigate the question addressed in \cite{friedman2016quantum,friedman2017quantum}, namely the problem of return to origin for a particle on an infinite lattice. We ask whether the exact results obtained by their  renewal approach   can be recovered using the effective non-Hermitian Hamiltonians discussed in \cite{dha15_pra,dha15_jpa}. We note that one can construct two effective non-Hermitian Hamiltonians, one with a small imaginary potential proportional to $\tau$, and  another with a large imaginary part proportional to $1/\tau$. Our main results here include analytic results, for the FDT problem, obtained from these non-Hermitian Hamiltonians and their comparison with earlier exact results. We also relate our results to those obtained  in \cite{mal14_jsp} on survival probability for non-Hermitian Hamiltonians.

As an application of the repeated measurement detection scheme, we use it for the so-called Aubry-Andr\'e-Harper model \cite{aubry1980analyticity,harper1955single} which is the case of a particle moving in a quasi-periodic potential.  This model has a transition from all states being localized to all states being extended as one changes the strength of the potential. We ask whether the different phases carry some signature regarding the form of the first detection distribution. 

The plan of the paper is as follows. In Sec.~\ref{sec:model} we define the precise model and dynamics composed of successive unitary evolutions followed by projections. We also write down the forms of the two effective non-
Hermitian Hamiltonians that are expected to describe the dynamics whenever the measurement time interval $\tau$ is small. In Sec.~\ref{sec:NH1} we consider the first non-Hermitian Hamiltonian with a small imaginary part. Using first order perturbation theory we find the eigenstates and eigenfunctions of the Hamiltonian and use them to compute the first passage time distribution and survival probability, and compare these with exact analytic and numerical results.  In Sec.~\ref{sec:NH2} we discuss the second non-Hermitian Hamiltonian and, use the approach
of \cite{mal14_jsp} to obtain some analytic results and again make comparisons with exact results. In Sec.~\ref{sec:AAH} we present results on survival probability in the Aubry-Andr\'e-Harper and finally we conclude with some discussions in Sec.~\ref{sec:conclusions}.

\section{Definition of model and mappings to non-Hermitian Hamiltonians}
\label{sec:model}

We consider a quantum particle on a one-dimensional lattice with  $N=2L+1$ sites. The system is described by the tight-binding Hamiltonian
\begin{align}
  H = -\gamma \sum_{x=-L}^{L-1} (|x\ra\la x+1| + |x+1\ra\la x|),
\end{align}
where $\gamma$ is the hopping strength which we set to the value one hereafter. 
We assume the  particle's wave-function is initially localized at a site $x=a$. 
Its free time-evolution would be described by   $|\psi(t)\ra = e^{-iH t}|a\ra$. However, the free evolution is interrupted by a sequence of projective measurements made, at regular time intervals $\tau$, to detect the presence of the particle at the origin $x=0$. The measurements continue till detection of the particle. Our main interest is in the probability $S_n$ that the particle survives detection up to the $n$th measurement and also the probability $p_n=S_{n-1}-S_{n}$ that it is detected for the first time on the $n$th measurement. 

Let us define the unitary evolution operator $U_\tau=e^{-iH \tau}$ and the projection operator $B=\hat I - |0\rangle \langle 0|$, corresponding to a projection of the system to a space complimentary to the detection site (the origion).  Let $|\psi_0\ra$ be the initially normalized wavefunction of the particle. 
It has been shown in \cite{dha15_pra,dha15_jpa} that the survival probability, \emph{after} the $n^{th}$ measurement. 
is simply given by $S_n=\langle \psi_n |\psi_n\rangle$, that is it can be expressed as the norm  of an appropriate  wave-function $|\psi_n\rangle$ with non-unitary  time evolution. 
The time evolution of $|\psi_n\rangle$, is precisely given by
\begin{align}
  |\psi_n \rangle &= \tilde{U} |\psi_{n-1}\rangle = \tilde{U}^n |\psi_{0}\rangle ,
\label{psievol} 
 \end{align}
 where $\tilde{U} = B U$ is the non-unitary evolution operator.

{\bf Mapping to non-Hermitian Hamiltonians} The quasi-Zeno dynamics refers to the case where $\tau$ is small but finite. In this limit, it was  shown in \cite{dha15_pra,dha15_jpa} that the dynamics can be described by an effective non-Hermitian Hamiltonian, thus one has
\begin{equation}
  \tilde{U}=e^{-i H_{eff} \tau}~.
\end{equation}
Two different non-Hermitian Hamiltonians were proposed and we now describe their forms for the present problem. \\
{\emph Mapping 1:}
According to the results derived in sec. III of \cite{dha15_pra}, the effective dynamics is on a lattice with $2L$ sites (excluding the measurement site), with the effective Hamiltonian given by
\begin{align}
  H_{eff}^{(1)} &= H_S^{(1)} + V_{eff}^{(1)};\label{heff1} \\
  H_S^{(1)} &= -\sum_{x=-L}^{-2} \left(|x\ra\la x+1| + |x+1\ra\la x|\right)\nn\\
                & \hspace{1cm}- \sum_{x=1}^{L-1} \left(|x\ra\la x+1| + |x+1\ra\la x|\right);\nn\\
  V_{eff}^{(1)} &= -\frac{i\tau}{2}\left(|1\ra\la 1| + |-1\ra\la -1| + |1\ra\la -1| + |-1\ra\la 1|\right). \nn
\end{align}
Note that $H_S^{(1)}$ is similar to the original Hamiltonian with the origin removed from the lattice. The presence of detector manifests itself through $V_{eff}^{(1)}$, where, for instance, the first term is the contribution to the effective Hamiltonian due to the particle hopping from site $x=1$ to the origin and back to $x=1$ on the original lattice.

{\emph Mapping 2:} A different mapping was also discussed in \cite{dha15_pra}, in which case the mapped system retains its original size of $2L+1$ sites, 
but now the imaginary potential consists of an on-site term of strength $-i2/\tau$ at the origin. The full Hamiltonian in this case is given by
\begin{align}
  H_{eff}^{(2)} &= H_S^{(2)} + V_{eff}^{(2)}; \label{heff2}\\
  H_S^{(2)} &= -\sum_{x=-L}^{L-1} (|x\ra\la x+1| + |x+1\ra\la x|);\nn\\
  V_{eff}^{(2)} &= -\frac{2i}{\tau}|0\ra\la 0|. \nn
\end{align}
In both cases the wave-function at time $t=n \tau$ is given by $|\psi(t)\rangle = e^{-i H_{eff} t} |\psi(0)\ra$. We note that for $L \to \infty$ the dynamics described by Eq.~\ref{heff2} is identical to that in \cite{mal14_jsp} with the identification of $2/\tau$ as the strength of the imaginary potential in that paper. This work obtained analytic results for the survival probability for $t\to \infty$ and we will discuss these, in Sec.~(\ref{sec:NH2}),  in the present context.

In the following two sections we will discuss specific predictions obtained for the detection probability, obtained from these two effective non-Hermitian Hamiltonian descriptions. We will also compare these predictions with exact numerical results for the measurement as well as analytic results from other approaches. In all numerical comparisions  we have chosen the value $\tau=0.1$.

\begin{figure}[!h]
  \centering
  \subfigure[]{\includegraphics[width=8cm]{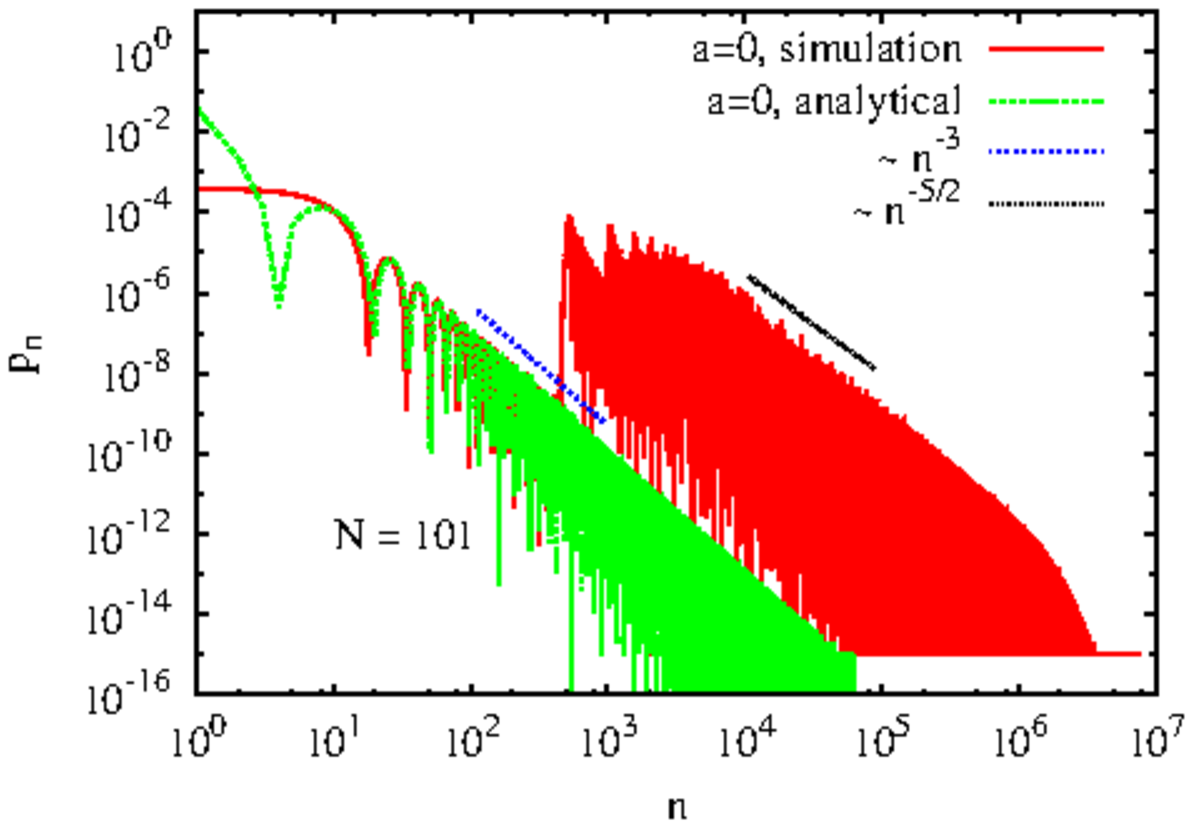}}
  \subfigure[]{\includegraphics[width=8cm]{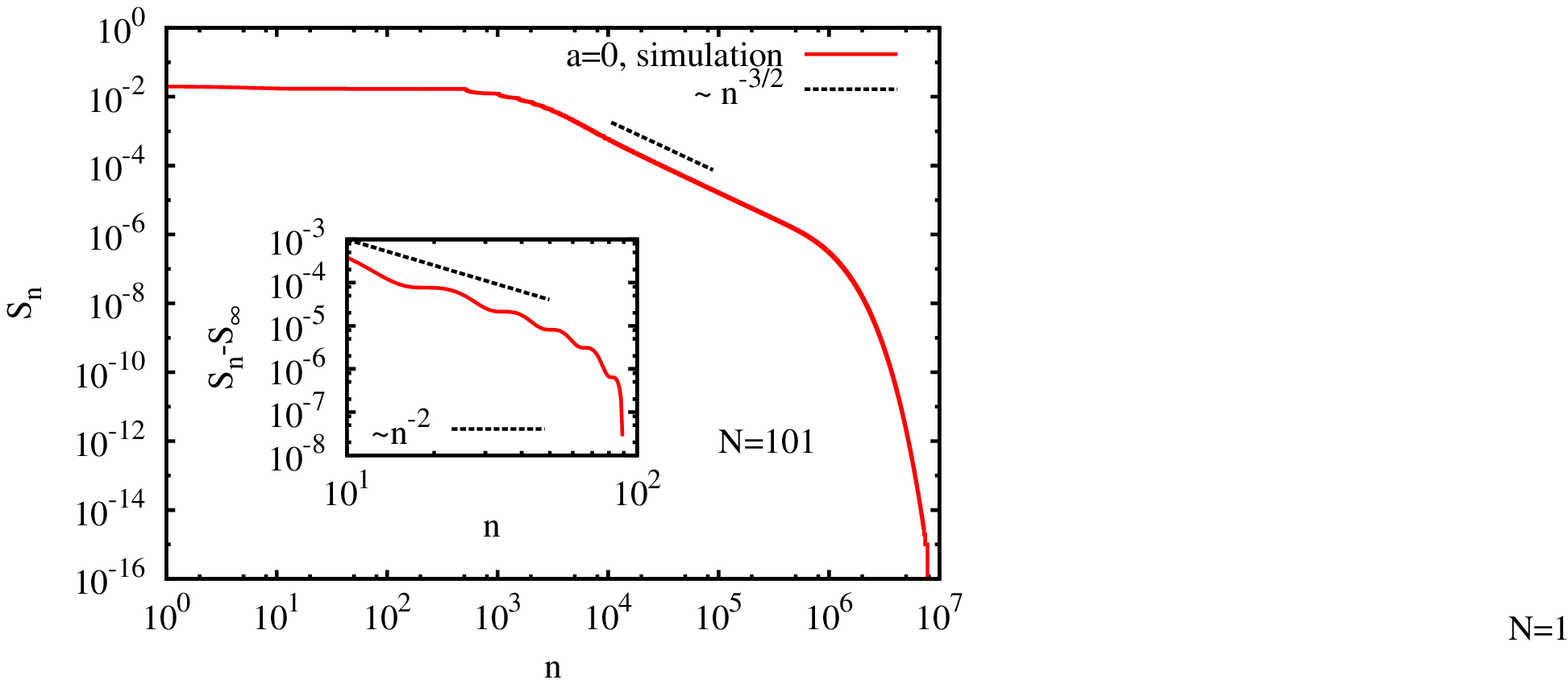}}
  \caption{(a) Plot of the first detection time distribution, obtained from simulation (red curve), compared to the analytical asymptotic form of the first detection time distribution [Eq.~(\ref{pn_barkai})] obtained in \cite{friedman2016quantum}. The particle is initially located at the origin $a=0$ and the detector is also at the origin. (b) Plot of survival probability $S_n$ as a function of $n$. The inset is a zoomed-in section of the full curve after subtraction of the ``saturation'' value $S_\infty = 0.01701$ .  The lattice size here is $N=101$ and detection time interval was set at $\tau=0.1$. }
  \label{fig:comparison_barkai} 
\end{figure}

{\bf{Finite size effects}}: 
For the infinite lattice, {\emph{i,e}} in the limit $L\to \infty$, the first return problem with this dynamics was studied in \cite{friedman2016quantum}, using a renewal equations approach. There it was shown that the first detection probability, in the asymptotic time limit, for a particle initially located at the origin ($a=0$), is given by
\begin{align}
  p_n \sim \frac{4\gamma \tau}{\pi n^3}\cos^2\bigg(2\gamma\tau n + \frac{\pi}{4}\bigg).
  \label{pn_barkai}
\end{align}
This result is valid for the case of an infinite lattice (i.e $N \to \infty$).  In Fig.~(\ref{fig:comparison_barkai}a), we show the comparison of the first detection time probabilities given by the above equation, with the result obtained from direct simulation on a lattice with $N=101$ sites. In this figure, the red solid line gives the FDT probability, obtained from exact numerics  using the relation $p_n = S_n-S_{n+1}$, with $S_n=\langle \psi_n |\psi_n\rangle$ and $|\psi_n\rangle$ given by  Eq.~(\ref{psievol}). The solid green  line corresponds to the analytical result given in  Eq.~\eqref{pn_barkai}.  
It is clear from this figure that the agreement with the analytical result holds only as long as  finite size effects do not show up ($n \lesssim 500$). In this region, the decay follows the $n^{-3}$ law. However, for values of $n$ much larger than about $500$, the decay in the exact numerics is observed to follow a $n^{-5/2}$ law. Finally, for very large values of $n$ ($\sim 10^6$), the decay becomes faster.
Thus, we see that one can identify three temporal regimes. For the example studied in   Fig.~(\ref{fig:comparison_barkai}), these three regimes  correspond to 
$0 << n \lesssim N,~ N \lesssim n \lesssim N^3,~$ and finally a regime with exponential decay  at times $n \gtrsim N^3$. The $O(N)$ time scale corresponds to ballistic propagation of perturbations in the system, while the $O(N^3)$ time scale can be understood as corresponding to the eigenvalue with the smallest imaginary part of a non-Hermitian Hamiltonian (see \cite{dha15_pra}). A second interesting point is in the behaviour of the survival probability in the different regimes, shown in Fig.~(\ref{fig:comparison_barkai}b). In the short-time regime the survival probability appears to saturate to the value $S_{n \to \infty}=0.01701...$. This means that on the infinite lattice, a particle starting from the origin will survive  detection with a finite probability (unlike the classical one-dimensional random walker). Subtracting this saturation value, we see the expected $1/n^2$ decay of the survival probability.  We will discuss this point further in later sections.

\section{Calculation of survival probability from $H^{(1)}_{eff}$ using first-order perturbation theory}
\label{sec:NH1}
In this approach, the evolution of the particle is given by $|\psi(t)\rangle = e^{-i H^{(1)}_{eff} t} | a\rangle$ where $H^{(1)}_{eff}$ is given by Eq.~(\ref{heff1}). The  survival probability is the norm of this state $S(t) =\langle \psi(t)|\psi(t) \rangle$. Let $\psi_r(x)$ and $\epsilon_r$ be the  right eigenvectors ($2L$ in number) and corresponding eigenvalues of $H^{(1)}_{eff}$. In general $\epsilon_r$ will have complex parts and we write $\epsilon_r=p_r-i q_r$. We assume orthogonality and completeness of the eigenfunctions though this will need to be verified in any specific example. We can then expand the initial state $|a\rangle $ in this basis as 
$|a\rangle = \sum_r \langle \psi_r |a\rangle | \psi_r \rangle$ and the time evolution 
is then given by $|\psi(t)\rangle = \sum_r e^{-i p_r t} e^{-q_r t} \langle \psi_r |a\rangle | \psi_r \rangle$. Hence the survival probability is given by
\begin{align}
  S(t)= \sum_r e^{-2 q_r t} |\langle \psi_r |a \rangle|^2~.
\end{align}
We will now proceed to obtain the eigenfunctions and eigenvalues of $H^{(1)}_{eff}$ by first order perturbation theory, by treating the imaginary potential as the perturbation. We first note that in the absence of the perturbation $V^{(1)}_{eff}$, the Hamiltonian consists of two disjoint Hermitian parts, with each part corresponding to a particle confined to either side of the origin.  
If the particle is to the left, let the state be denoted by $|\phi^L\ra$. If it is on the right, its state is denoted by $|\phi^R\ra$. The $k^{th}$ eigenfunction of the particle, corresponding to energy $e_k=-2\cos[k \pi/(L+1)]$, is
\begin{align}
  \phi^L_k(x) &=
                \begin{cases}
                  \sqrt{\frac{2}{L+1}}\sin\left(\frac{k\pi x}{L+1}\right), \hspace{0cm}(x=-L,-L+1,\ldots,-1)\\
                  0, \hspace{2.5cm} (x=1,2,\ldots,L)
                \end{cases}\nn\\
  \phi^R_k(x) &=
                \begin{cases}
                  0, \hspace{2.5cm}(x=-L,-L+1,\ldots,-1)\\
                  \sqrt{\frac{2}{L+1}}\sin\left(\frac{k\pi x}{L+1}\right), \hspace{0cm}(x=1,2,\ldots,L).
                \end{cases}
  \label{phi_ab}
\end{align}
\begin{figure}[!h]
  \centering
  \subfigure[]{\includegraphics[width=8cm]{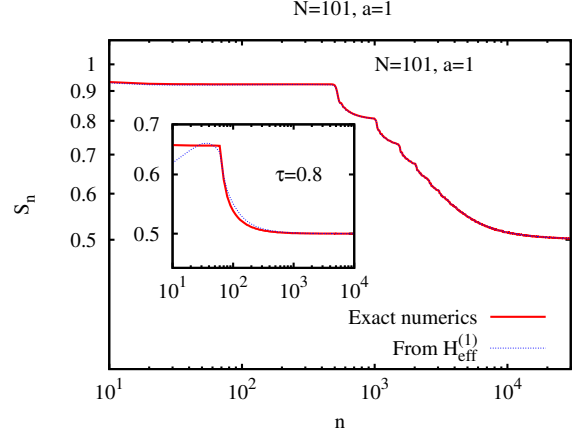}}
  \subfigure[]{\includegraphics[width=8cm]{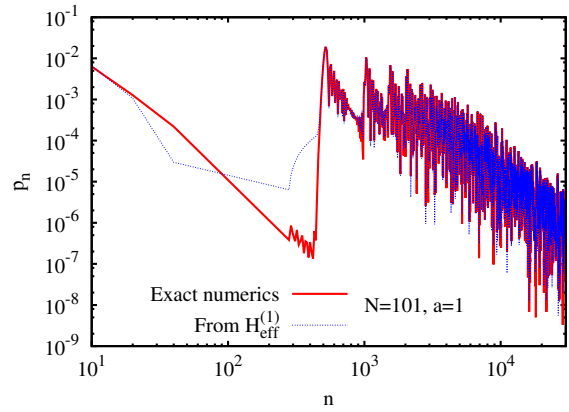}}
  \caption{(a) Survival probability plotted as a function of time for a particle starting at site $a=1$, computed by exact numerical evolution under projective measurements (red curve) and by computing it using Eq.~\eqref{full_wf1} (green curve). The meaurement time interval was $\tau=0.1$. \textit{Inset:} Comparison of the two plots  for  larger time interval of   $\tau=0.8$. In this case we clearly observe the discrepancy between them. 
    (b) Plot of the  first detection probability obtained from the two data sets in (a).}
  \label{fig:surv_prob_direct}
\end{figure}
Alternatively we can construct, for each $k = 1,2,\ldots,L$, two degenerate eigenfunctions of $H_S$ as follows:
\begin{align}
  \phi^+_k(x) &=  \frac{1}{\sqrt 2}\left[ \phi^L_k(x) + \phi^R_k(x) \right] \nn \\
  \phi^-_k(x) &=  \frac{1}{\sqrt 2}\left[ \phi^L_k(x) - \phi^R_k(x) \right]~,\nn\\
              &~~{\rm for }~x=-L,-L+1,\ldots,-1,1,2,\ldots,L~.
                \label{phi+-}
\end{align}
Now, operating with $H_{eff}$ on these eigenfunctions, we observe that  the set of states $|\phi_k^+\ra$ continue to be exact eigenstates. Thus we set
\begin{align}
  \psi_k(x)\equiv\psi^+_k(x) =\phi^+_k~,~~~\epsilon_k=e_k,~~{\rm for}~ k=1,2,\ldots,L~.
\end{align}
We find the remaining eigenstates $\psi_k$ $(k=L+1,L+2,\ldots, 2 L)$ by perturbation of the states $\phi^-_k$. 
From standard first-order perturbation theory, we get 
\begin{align}
  \psi_{L+k}(x)&\equiv \psi^-_k(x) \nn\\
               &=\phi^-_k(x)+  \sum_{k' \neq k } \left(\frac{\la \phi^-_{k'}|V_{eff}|\phi^-_k\ra}{e_k-e_{k'}}\right)\phi^-_{k'}(x)   \label{phi1a} \\
  \epsilon_{L+k} &= e_k+ \la\phi^-_k|V_{eff}|\phi^-_k\ra \nn \\
               &= e_k-i \beta_k, ~~~{\rm with}~ \beta_k =  \frac{2 \tau}{L+1}\sin^2\left(\frac{k\pi}{L+1}\right)~.          \label{beta} 
\end{align}
The time evolution of the wave function can thus be written as
\begin{align}
  \psi(x,t) &= \sum_{k=1}^L e^{-i e_k t} \left[ {\psi^+_k}(a) \psi^+_k (x)\right.\nn\\
            &\hspace{1cm}\left.+ e^{-\beta_k t} {\psi^-_k}(a) \psi^-_k(x) \right]~.
              \label{full_wf1}
\end{align}
We note that this has been obtained at order $O(\tau)$ and their orthonormality (with appropriate left eignevectors) is valid up to corrections  $O(\tau^2)$. The norm of the above wave-function gives the survival probability. 
%
%
We tested the analytical prediction from Eq.~\eqref{full_wf1}) against exact numerics done on a finite ($N=101$) lattice. The results, for $S_n$ and $p_n$, are shown in Figs.~(\ref{fig:surv_prob_direct}a,\ref{fig:surv_prob_direct}b)  where we see that the agreement between data from the exact numerics and results from the non-Hermitian Hamiltonian are almost in perfect agreement at all times. The saturation value $0.5$ corresponds to the fact that at long times the contribution of the states $\psi_n^-$ decays to zero while the remaining part has a normalization $1/2$. We can see another plateau at times order $\lesssim N$ and this corresponds to the value of the survival probability of a particle on the  infinite lattice. This will be
discussed in more detail in the next section. In the inset of Fig.~\ref{fig:surv_prob_direct}a we compare the values of $S_n$ obtained from direct simulation and perturbation theory, when $\tau=0.8$. The disagreement is expected on two grounds. Firstly, a large value of $\tau$ breaks the mapping between the actual evolution and the evolution under $H_{eff}^{(1)}$. Secondly, it invalidates the justification for the usage of perturbation theory in this case.


\section{Calculation of survival probability from $H^{(2)}_{eff}$}
\label{sec:NH2}
\begin{figure}[!h]
  \centering
  \subfigure[]{\includegraphics[width=8cm]{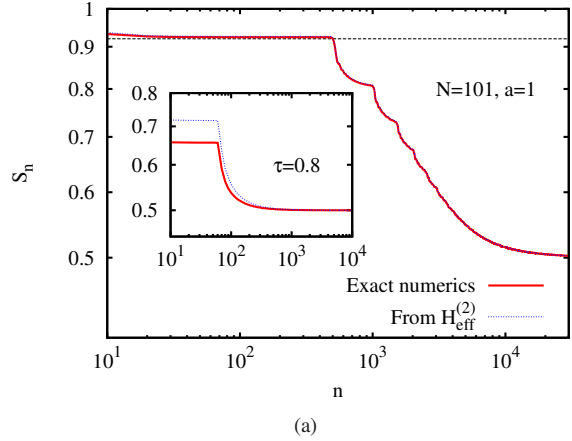}}
  \subfigure[]{\includegraphics[width=8cm]{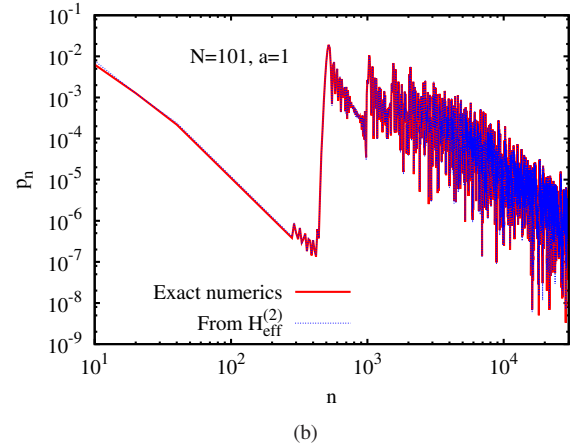}}
  \caption{(a) Comparison of survival probabilities for exact numerics and with evolution under $H_{eff}^{(2)}$, for $\tau=0.1$. The dashed line indicates the value of $S_{\infty}$ for an infinite lattice. \textit{Inset:} The inset shows the disagreement between the two curves for $\tau=0.8$. (b) Comparison of the corresponding first detection times.}
  \label{fig:mapping2}
\end{figure}

As discussed in sec.~\ref{sec:model}, the evolution under projective measurements admits a second  mapping to a system where there is a large imaginary potential at the measurement sites, in addition to the regular hopping Hamiltonian. In Figs.~(\ref{fig:mapping2}a,\ref{fig:mapping2}b), we show numerical results for the case of a finite lattice ($N=101$), which clearly shows that the  mapping between the projective dynamics and this second non-Hermitian dynamics holds quite accurately for  the value $\tau = 0.1$~. In the inset of Fig.~\ref{fig:mapping2}a, we have shown that the mapping of exact evolution with the evolution under $H_{eff}^{(2)}$ when $\tau=0.8$ is large enough.

 For our 1D lattice problem with measurements at the origin, the non-Hermitian dynamics  is identical to the one discussed in \cite{mal14_jsp}, where exact results were obtained for the survival probability on the infinite lattice.
For the case of the infinite lattice, we now outline the approach of \cite{mal14_jsp} and show that it leads to reproduction of the results in \cite{friedman2016quantum}. 
For the infinite lattice we note that the  evolution equation for the wave function is given by $i\frac{d}{dt}|\psi\ra = H_{eff}^{(2)}|\psi\ra$, where
\begin{align}
  H_{eff}^{(2)} = -\sum_{x=-\infty}^{\infty}(|x+1\ra\la x| + |x\ra \la x+1|) - i\Gamma|0\ra\la 0|, 
\end{align}
with $\Gamma = 2/\tau$.
The corresponding Schr\"odinger equation is
\begin{align}
  i\pd{\psi_x}{t} = -\psi_{x+1}(t) - \psi_{x-1}(t) - i\Gamma \delta_{x0}\psi_0(t)~.
  \label{schrod}
\end{align}
We solve this equation for a localized initial condition  $|\psi(t=0)\ra=|a\ra$ to  get the wave-function $|\psi(t)\ra$ at any time $t$.
Let us define the  Laplace transform of the wave-function $\psi_x(t)$  as
\begin{align}
  \tilde\psi_x(s) &= \int_0^\infty dt ~\psi_x(t) e^{-st}.
\end{align}
and the  Fourier-Laplace transform as
\begin{align}
  \hat\psi(q,s) &= \sum_{x=-\infty}^{+\infty}\int_0^\infty dt ~\psi_x(t) e^{iqx-st}.
\end{align}
From the equation of motion we then get%
\begin{align}
  \hat\psi(q,s) &= \frac{i(e^{-iqa}-\Gamma\tilde\psi_0(s))}{is-2\cos q},
                  \label{fourier-laplace}
\end{align}
where the term $\tilde{\psi}_0(s) =\int_0^\infty dt e^{-st} \psi_0(t)=[1/(2\pi)] \int_0^{2 \pi} dq \hat{\psi}(q,s)$ can be immediately obtained as
\begin{align}
  \tilde\psi_0(s) = \frac{1}{\Gamma+\sqrt{s^2+4}}\left(\frac{\sqrt{s^2+4}-s}{2i}\right)^a.
\end{align}
Given the wave-function, the survival probability is simply given by the norm $\la \psi(t)|\psi(t) \ra$. Alternatively we note that the decay of the probability is caused by absorption at the origin, and it is easy to see therefore that the first detection probability density is given by 
\begin{equation}
  p(t)=-dS(t)/dt=2 \Gamma |\psi_0(t)|^2~.
\end{equation}
Hence one can obtain $S(t)=1-2 \Gamma \int_0^t |\psi_0(t)|^2$.

The main focus in 
\cite{mal14_jsp} was in computation of the survival probability at infinite times and analytic exact expressions were obtained for $S(\infty)$ for different initial conditions $x=a$. However the paper also obtains the exact form of $\psi_0(t)$ for large $\Gamma$ and from this one gets (for $a \neq 0$)
\begin{equation}
  p^{(a)}(t)=2 \frac{a^2}{\Gamma} \frac{J^2_a(2 t)}{t^2} \sim \left(\frac{\tau a^2}{\pi}\right) \frac{\cos^2(2 t -a\pi/2-\pi/4)}{t^{3}}~,
  \label{pt}
\end{equation}
where we used the asymptotic time expansion  of the Bessel function.
The superscript in $p^{(a)}$ denotes the initial position $a$.

For $a=0$, we can first evolve the system for one time step and then use the earlier result. We note that at the end of the first projective measurement, the evolution under $H_{eff}^{(2)}$ leads to the state
\begin{align}
  |\psi(\tau^+)\rangle &= -i\tau(|1\rangle + |-1\rangle),
\end{align}
up to first order in $\tau$. This can then evolved in time to obtain the final state
\begin{align}
  |\psi(t)\rangle &= -i\tau e^{-iH_{eff}^{(2)}(t-\tau)}(|1\rangle + |-1\rangle).
\end{align}
Making use of the symmetry between the states $|1\rangle$ and $|-1\rangle$, we can write
\begin{align}
  \psi_0(t) &= \la 0|\psi(t)\ra = -2i\tau \la 0|e^{-iH_{eff}^{(2)}(t-\tau)}|1\ra.
\end{align}
Thus the first detection probability at time $t$ is given by
\begin{align}
  p^{(0)}(t) &= |\psi_0(t)|^2 = 4\tau^2 p^{(1)}(t-\tau)\approx 4\tau^2 p^1(t),
\end{align}
up to leading order in $\tau$. 
From Eq. \eqref{pt}, we therefore find that the first detection probability is
\begin{align}
  p^{(0)}(t) &= \left(\frac{4\tau^3}{\pi}\right) \frac{\cos^2(2 t +\pi/4)}{t^{3}}.
\end{align}
This agrees with Eq.~(\eqref{pn_barkai}) when we note that continuous time $t=n\tau$ and $p_n = \tau p^{(0)}(t)$. Thus, we obtain  agreement between the results obtained using  \cite{mal14_jsp} and those in \cite{friedman2017quantum}.

\begin{figure}[!h]
  \centering
  \subfigure[]{\includegraphics[width=8cm]{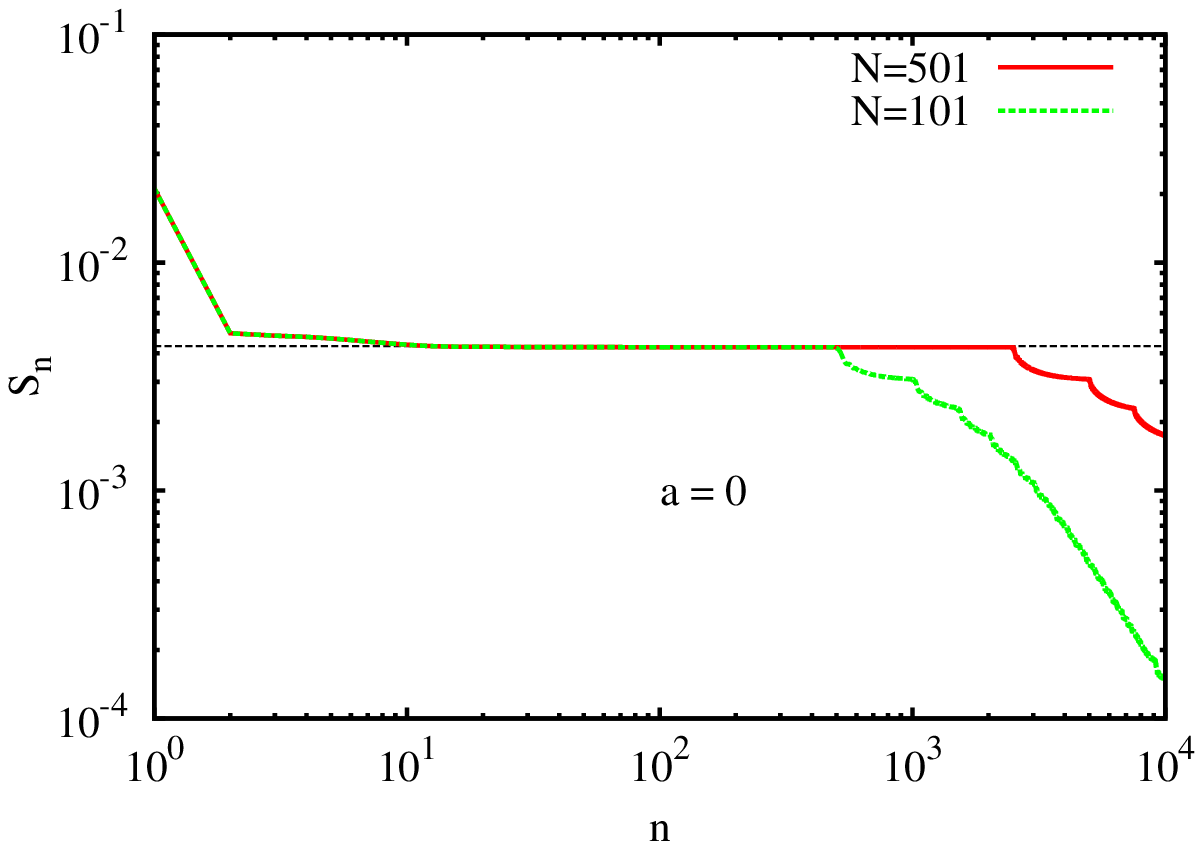}}
  \subfigure[]{\includegraphics[width=8cm]{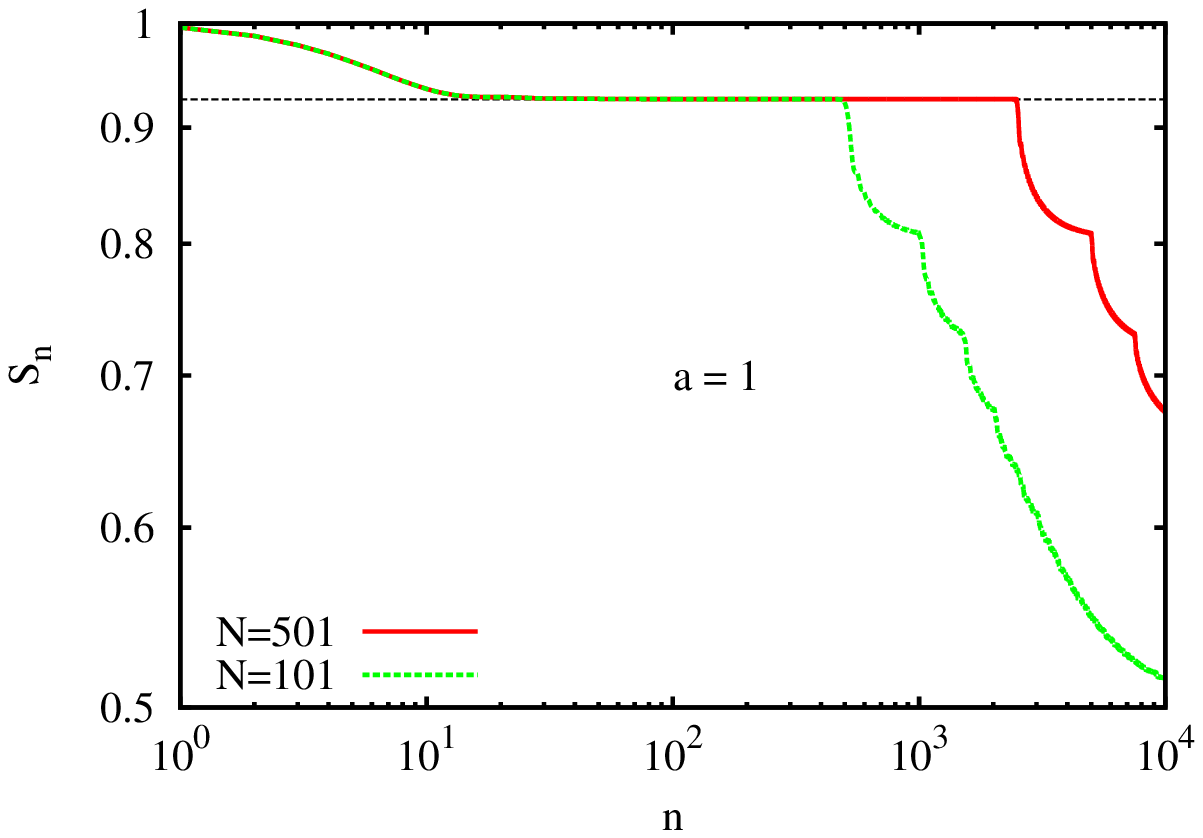}}
  \caption{Variation of survival probability with time, when the particle is initially located at the origin (a) and at the site next to the origin (b), for two sample sizes: $N=101$ and $501$. The plots are from exact numerical evolution using the non-Hermitian Hamiltonian $H_{eff}^{(2)}$. The horizontal dashed lines indicate the values $0.00125$, for $a=0$,  and  $0.92$, for 
$a=1$, of $S_{\infty}$ obtained from the analytic result for an infinite lattice.}
  \label{fig:surv_prob_centre}
\end{figure}

The authors in \cite{mal14_jsp} were also able to obtain the form of $S_\infty$.
Interestingly, it is seen  that, for all initial starting positions,  the survival probability is finite and (except for the case $a=0$) goes to unity as $\Gamma \to \infty$ --- this corresponds to the Zeno limit. For the special case,  
$a=0$,  $S_\infty$ vanishes as $1/\Gamma^2$. 
As discussed in Sec.~(\ref{sec:model}), on a finite lattice there are three temporal regimes. For times $t \lesssim N$, the results are  independent of system size and we basically see results as in the infinite lattice.  
We find that $S_n$ initially decays with $n$ and then shows a plateau region, beyond which ($n \gtrsim N$), finite size effects begin to show up and the survival probability eventually decays to a value $1/2$ (for $a>0$) or to zero (for $a=0$). It is the initial plateau region that corresponds to the value of $S_\infty$ for an infinite lattice. To verify these facts we have plotted, in Figs.~\ref{fig:surv_prob_centre}(a,b),  $S_n$ as a function of $n$ for two different lattice sizes $N=101$ and $501$, for the initial positions $a=0$ and $a=1$ respectively. We clearly see that the initial plateau region survives for a longer time (scaling as $\sim N$) for the larger lattice. The plots have been generated by evolving the system with the Hamiltonian $H^{(2)}_{eff}$. In the limit of an infinite lattice, there would be no decay of this plateau at all.
We  also compare these plateau values with the analytical values of $S_{\infty}$ for the same initial positions, as obtained in \cite{mal14_jsp}. We observe that the agreement between the two approaches is excellent. 

\section{First detection time distribution in the Aubry-Andr\'e-Harper model}
\label{sec:AAH}

\begin{figure}[!h]
  \centering
  \subfigure[]{\includegraphics[width=8cm]{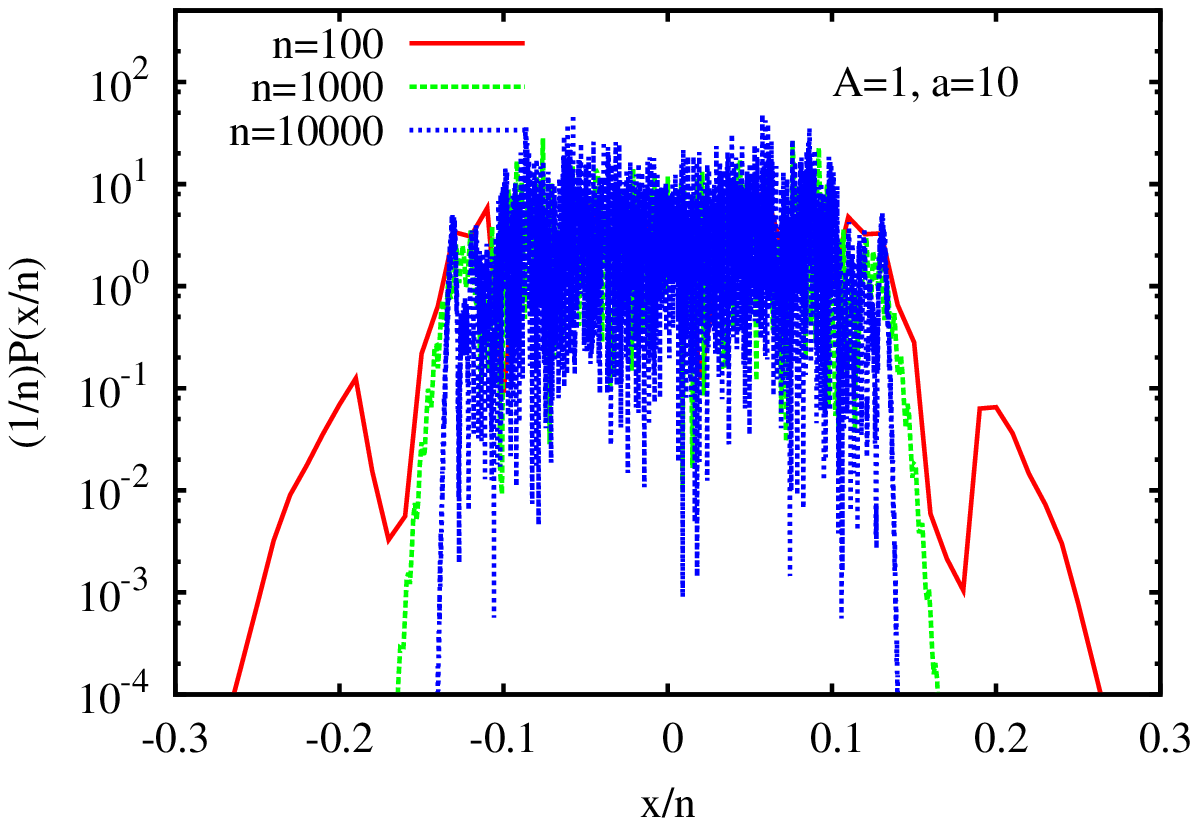}}
  \subfigure[]{\includegraphics[width=8cm]{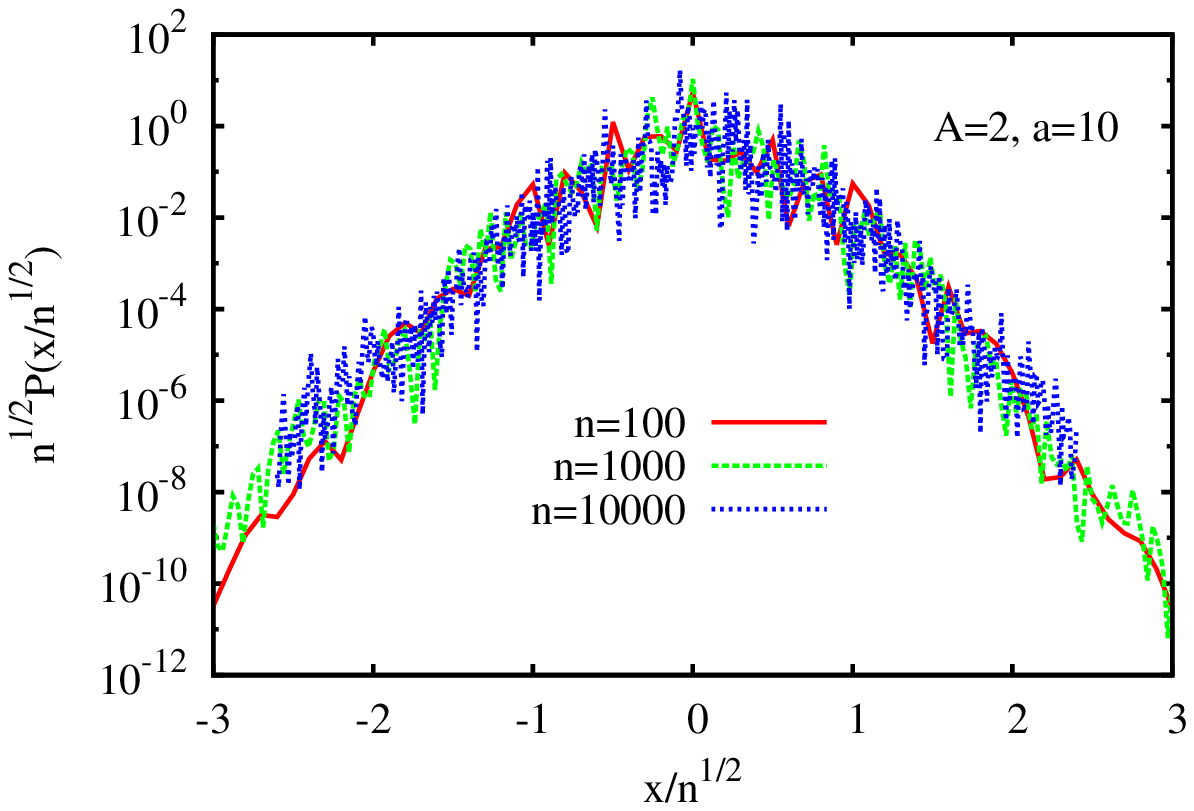}}
  \subfigure[]{\includegraphics[width=8cm]{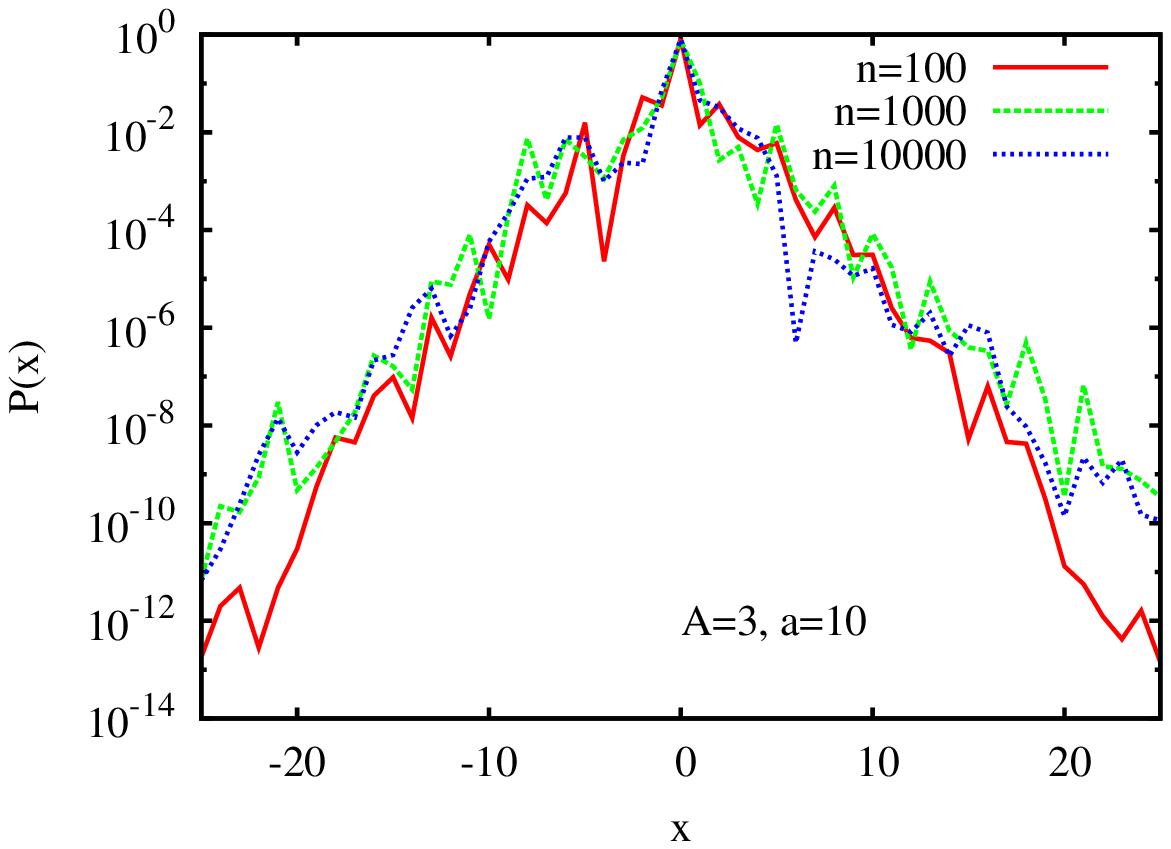}}
  \caption{Scaled probability distributions for position for (a) $A=1$, (b) $A=2$ and (c) $A=3$. The lattice size for $A=1$ is $N=5001$. For $A=2$ and $A=3$, it is $N=501$.}
  \label{fig:spreading_AA}
\end{figure}

In the AAH model \cite{aubry1980analyticity,harper1955single} one considers the
motion of a quantum particle in a quasiperiodic potential. The Hamiltonian is given by 
\begin{align}
  H &= -\sum_{x=-L}^{L-1} (|x\ra\la x+1| + |x+1\ra\la x|) \nn \\
    &+ \sum_{x=-L}^{L1} A\cos(2\pi x\sigma) |x\ra \la x|,
\end{align}
where $\sigma$ is an irrational number, which we here choose to be the golden ratio $\sigma = (\sqrt{5} + 1)/2$. This system shows a remarkable transition in properties of eigenstates as one changes the amplitude $A$ of the potential.
For $A<2$, all eigenstates are delocalized, whereas for $A>2$ all states are localized. At the critical point $A=2$, the eigenstates are neither localized nor extended and wave-packets show diffusive spreading. 
We illustrate the differences in the three phases in Fig.~\ref{fig:spreading_AA},  where the probability of the particle being at site $x$, $|\psi_x(t)|^2$, given that it starts from the center of the lattice, has been plotted as a function of $x$ at different times $t=n\tau$ with $\tau=0.1$. These are for the free evolution under $H$, in the absence of any detector. The red solid line, green dashed line and blue dotted line show the probability distributions at times $n=100$, $1000$ and $10000$ respectively, and the three data sets have been scaled according to their expected behavior. In particular, we note that for $A=2$ the probability density spreads diffusively, though it has recently been noted that the wave-form is non-Gaussian \cite{purkayastha2017sub}. 

A natural question of interest is to ask whether one can see a marked difference in the form of the survival probability (under our measurement dynamics) in the three different phases. For example, would the survival probability decay to zero in the long time limit, in the critical phase?  
We now address this question, in particular the time-dependence of $S_n$ and $p_n$ in the long-time limit in a thermodynamic system (i.e after taking $L \to \infty$).

\begin{figure}[!h]
  \centering
  \subfigure[]{\includegraphics[width=8cm]{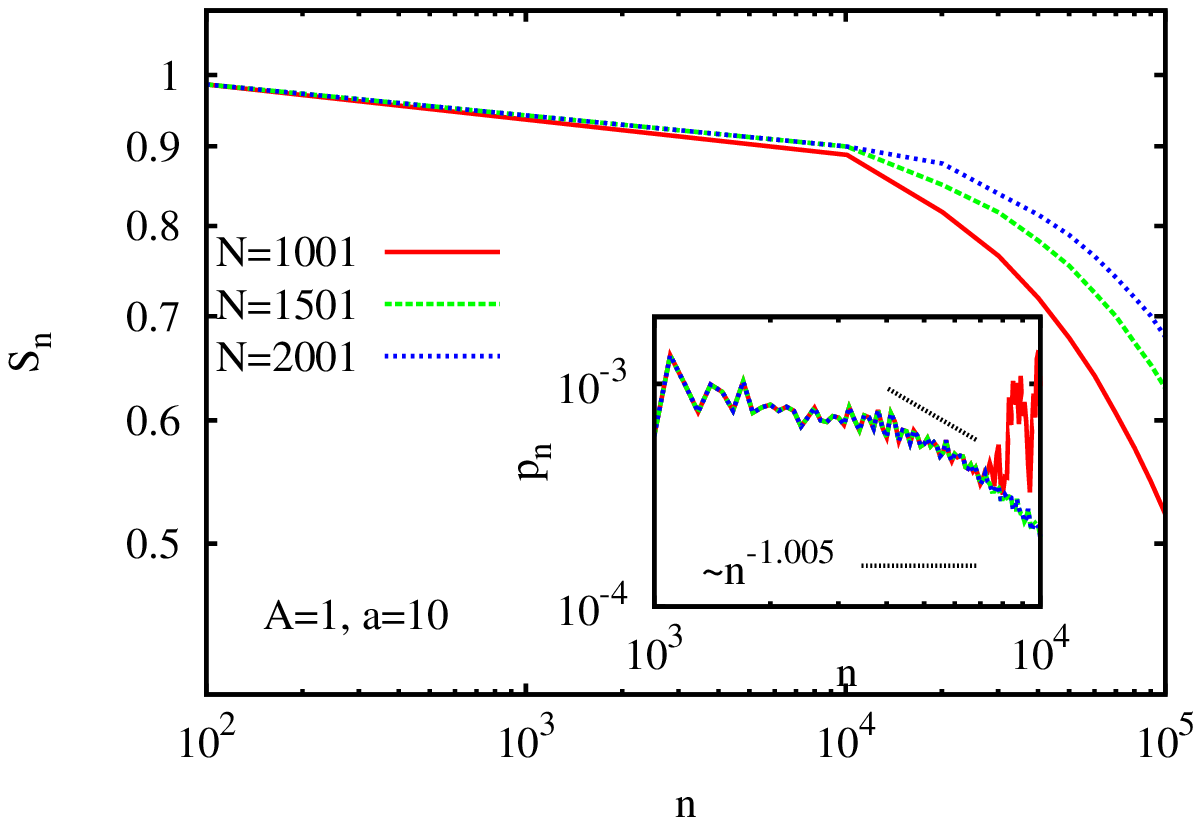}}
  \subfigure[]{\includegraphics[width=8cm]{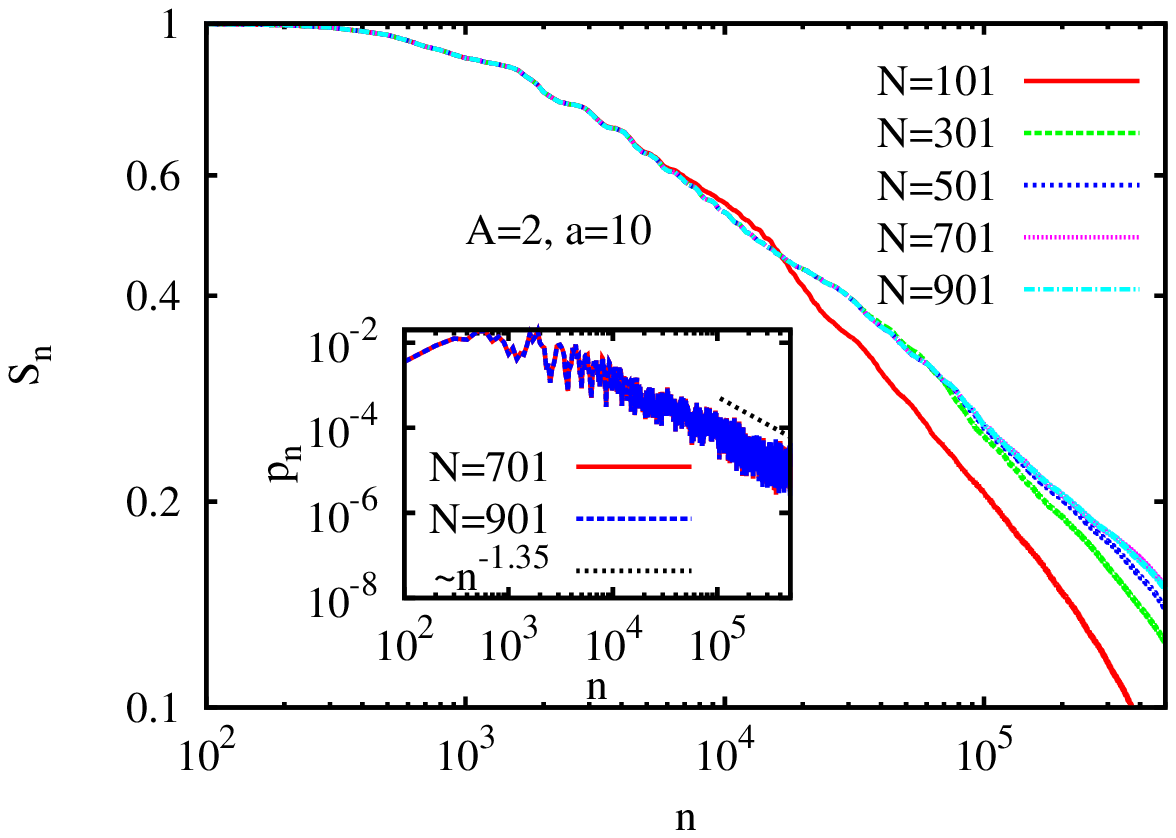}}
  \subfigure[]{\includegraphics[width=8cm]{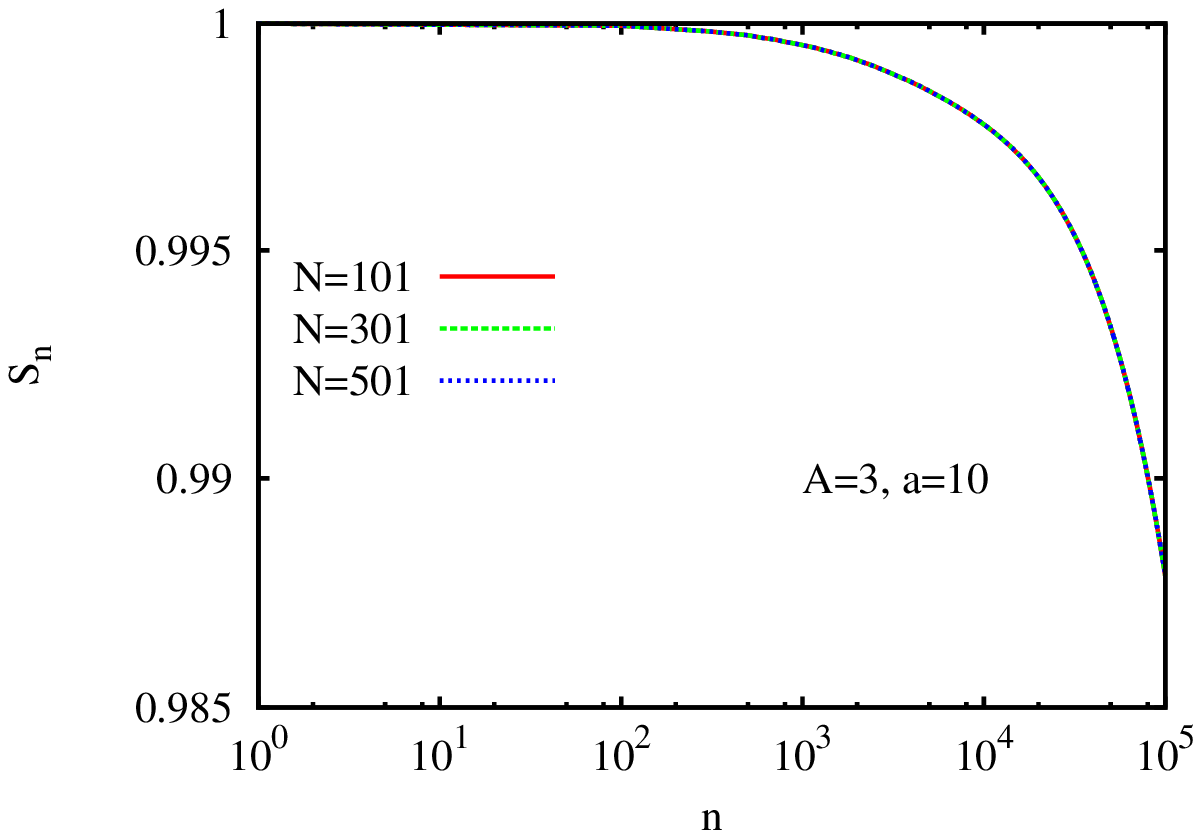}}
  \caption{Plot of $S_n$ with $n$ in the AAH model for different lattice sizes, for  (a) $A=1$, (b) $A=2$ and (c) $A=3$. The insets in (a) and (b) show the corresponding first detection time probabilities $P_n$ and the power law decays indicate the behavior of the temporal regime corresponding to an infinite lattice.}
  \label{fig:AA}
\end{figure}

In Figs.~\ref{fig:AA}, we have plotted these $S_n$ as a function of the number of measurements $n$ for the amplitude values $A = 1$, $2$ and $3$. We obtained data for lattices of different sizes which are prepared such that the central overlapping region for two different system sizes see exactly the same quasi-periodic potential.  The initial condition is always chosen to be a site that is $10$ lattice sites from the origin where detection occurs. The temporal region where the $S_n$ data for the different system sizes overlap, corresponds effectively to the infinite size limit. Hence we identify this time-scale and look at the decay of $P_n$ in this region.  As expected we find that this time scale is $O(N)$ for $A=1$, is $O(N^2)$ for $A=2$ and $O(1)$ for $A=3$. The main conclusions that we can draw from these plots are:

(i) For $A=1$ corresponding to ballistic spreading of wave-packet, we find a decay of $p_n$ slightly faster than $1/n$. This indicates a decay $S_n \sim 1/\log{n}$. We contrast this with the purely ordered case ($A=0$) where there the survival probability   saturates to a constant value [see e.g Fig.~(\ref{fig:surv_prob_centre})]. 

(ii) At the critical point, $A=2$, we see a faster decay $p_n \sim 1/n^{1.35}$ which implies again a decay of the survival probability $S_n \sim 1/n^{0.35}$. However the power seems to depend on initial conditions. 

(iii) Finally in the localized phase, $A=3$, the survival probability almost shows no decay. 


\section{Conclusions}
\label{sec:conclusions}

The first detection time distribution and survival probability for a quantum particle were studied on  finite and infinite one-dimensional open lattices, with a detector at the middle site. The measurement process consists of a sequence of 
selective projections,  made at regular time intervals $\tau$ between the particle's unitary evolution, till the particle is detected. We first showed, from  numerical computations, that for small $\tau$, results from the  exact dynamics were quite accurately reproduced by two different effective non-Hermitian Hamiltonians ($H_{eff}^{(1)}$ and $H_{eff}^{(2)}$). We showed that for the case of $H_{eff}^{(1)}$ which consisted of a small imaginary part, one could obtain analytic results 
for the survival probability using perturbation theory. On the other hand, for the case of $H_{eff}^{(2)}$, which has a large imaginary part, one can use existing results for the infinite lattice to obtain results for the $S_n$ which agree 
with results obtained by other approaches. 

We also pointed out that, for the finite lattice (with no on-site potentials), there are three time-scales (i) a time scale $\lesssim O(N)$ where the behavior of $S_n$ corresponds to $N\to \infty $ limit and we get $P_n \sim 1/n^3$,  (ii) a time scale $N \lesssim n \lesssim N^3$ where we find $P_n \sim 1/n^{5/2}$ and (iii) the large time limit  $N^3 \lesssim n$, where $P_n \sim e^{-n/N^3}$. 
For the infinite lattice, the survival probability saturates to a finite value (initial condition dependent) as $S_n \sim S_\infty - A/n^3$~, where $S_\infty$ is known exactly using the mapping to $H_{eff}^{(2)}$ (and results from \cite{mal14_jsp}).

Finally we studied the survival probability for a quantum particle in the Aubry-Andr\'e-Harper model, which has extended, localized  and critical phases depending on the amplitude of the quasi-periodic potential. We found numerical evidence that here, the survival probability for the infinite lattice decays to zero in 
the extended phase as $\sim 1/\log(n)$ and in the critical phase as $\sim 1/n^{0.35}$.    There has been increased interest in understanding the dynamics of projective measurements in experimental systems \cite{semba15,siddiqui16,huard15,murch} and experimental probes to address the aspects addressed in this paper should be possible in the near future.

\section{acknowledgment}
AD would like to thank support from the Indo-Israel joint research project No. 6-8/2014(IC), from the grant EDNHS ANR-14-CE25-0011 of the French National Research Agency (ANR) and from Indo-French Centre for the Promotion of Advanced Research (IFCPAR) under project 5604-2. AD thanks Subinay Dasgupta for helpful discussions.

%


\end{document}